    \documentclass[twocolumn,final]{svjour3}
\smartqed  % flush right qed marks, e.g. at end of proof
\usepackage{graphicx}
 \usepackage{mathptmx}      % use Times fonts if available on your TeX system
    \usepackage{bm}

    \usepackage{dcolumn}

    \usepackage{amsmath}

    \usepackage{amssymb}

    \usepackage[usenames]{color}

\newcommand\beq{\begin{equation}}
\newcommand\eeq{\end{equation}}
\newcommand\beqa{\begin{eqnarray}}
\newcommand\eeqa{\end{eqnarray}}
\newcommand{\nn}{\nonumber\\}

\newcommand{\s}{\infty}

\newcommand{\dd}{{\text{d}}}

\newcommand{\al}{\alpha}

\newcommand{\atwop}{\left(\frac{1+\alpha}{2}\right)}
\newcommand{\sg}{(\widehat{\bm{\sigma}}\cdot\mathbf{g})}
\newcommand{\sv}{(\widehat{\bm{\sigma}}\cdot\mathbf{V}_1)}
\newcommand{\svv}{(\widehat{\bm{\sigma}}\cdot\mathbf{V}_2)}

\newcommand{\nutz}{\nu_{2|0}}
\newcommand{\nuzt}{\nu_{0|2}}

\newcommand{\omeganu}{\nu}

 \journalname{Granular Matter}
\begin{document}

\title{Collisional rates for the inelastic Maxwell model: application to the divergence of anisotropic high-order velocity moments in the homogeneous cooling state
}

\titlerunning{Collisional rates for the inelastic Maxwell model}        % if too long for running head

\author{Andr\'es Santos         \and
        Vicente Garz\'o
}


\institute{A. Santos \and V. Garz\'o \at
              Departamento de F\'{\i}sica, Universidad
de Extremadura, E-06071 Badajoz, Spain \\
                           \email{andres@unex.es, vicenteg@unex.es} }

\date{\today}

\maketitle

\begin{abstract}
The collisional rates  associated with  the isotropic velocity moments $\langle
V^{2r}\rangle$ and the anisotropic moments $\langle V^{2r}V_i\rangle$ and $\langle
V^{2r}(V_iV_j-d^{-1}V^2\delta_{ij})\rangle$ are exactly derived in the case
of the inelastic Maxwell model as functions of the exponent $r$, the
coefficient of restitution $\alpha$, and the dimensionality $d$. The
results are applied to the evolution of the moments in the homogeneous free
cooling state. It is found that, at a given value of $\alpha$, not only the isotropic moments of a degree higher than a certain value diverge but also  the anisotropic moments do. {This implies that, while the \emph{scaled} distribution function has been proven in the literature to  converge  to the \emph{isotropic} {self-similar} solution in well-defined mathematical terms, nonzero initial anisotropic moments do not decay with time. On the other hand, our results show that} the ratio between an anisotropic moment and the isotropic moment of the same degree tends to zero.
\keywords{Inelastic Maxwell model \and Collisional rates \and Homogeneous cooling state}
\end{abstract}

\section{Introduction}

The prototypical model of a granular gas consists of a system of (smooth) inelastic hard spheres (IHS) with a constant coefficient of normal restitution $0<\alpha\leq 1$ \cite{BP04}. Under low-density conditions, the one-particle velocity distribution function $f(\mathbf{r},\mathbf{v};t)$ obeys the (inelastic) Boltzmann equation.
On the other hand, because of the intricacy of the collision operator, one has to resort to approximate or numerical methods to get explicit results, even in the elastic case ($\alpha=1$). The main mathematical difficulty lies in the fact that the collision frequency of IHS is proportional to the relative velocity of the two colliding particles.
As in the elastic case \cite{E81,TM80}, a significant way of overcoming the above problem is to apply a mean-field approach whereby the collision frequency is replaced by an effective quantity independent of the relative velocity. This defines the so-called inelastic Maxwell model (IMM), which has received much attention in the last few years, {especially in the applied mathematics literature (see, for instance, \cite{BMP02,BK00,BK02,BCT06,BCG00,BC03,BCG09,BCT03,BG06,BC07,BGM10,CCC09,CCG00,EB02a,EB02b,FPTT10,GS07,KB02,S03}} and the review papers {\cite{BK03,BCG08,BE04,CT07,GS11}}).

Although the Boltzmann equation for the IMM keeps being a mathematically involved nonlinear integro-differential equation, a number of exact results can still be obtained. In particular, the collisional velocity moments of a certain degree $k$ can be exactly expressed as a bilinear combination of velocity moments of degrees $k'\leq k$ and $k''=k-k'$. Of course, the terms with $k'=k$ or $k''=k$ are products of a moment of degree $k$ and a coefficient proportional to density (moment of zeroth degree). We will refer to the latter coefficient as a \emph{collisional rate}. While all the collisional rates have been evaluated in the one-dimensional case \cite{BK00}, to the best of our knowledge, only the ones related to the isotropic moments  of any degree {\cite{BK03,EB02b}} and those related to isotropic and anisotropic moments of degree equal to or smaller than four \cite{GS07} have been obtained for general dimensionality $d$.

The aim of this paper is to derive the collisional rates associated, not only  with  the isotropic velocity moments $\langle
V^{2r}\rangle$, but also  with the anisotropic moments $\langle V^{2r}V_i\rangle$ and $\langle
V^{2r}(V_iV_j-d^{-1}V^2\delta_{ij})\rangle$. This is done by a method alternative to that followed in {Refs.\ \cite{BK03,EB02b}} for the isotropic moments. The knowledge of the above collisional rates is applied to the study of the time evolution of the moments in the homogeneous cooling state (HCS).
It is known that the isotropic moments, scaled with respect to the thermal velocity, diverge in time beyond a certain degree that depends on $\alpha$,  as a consequence of the algebraic high-velocity tail exhibited by the HCS {self-similar} solution \cite{BK02,EB02a,EB02b}.
The relevant finding of our study is that, at a given value of $\alpha$,  also the \emph{anisotropic} moments diverge beyond a certain degree. {This might seem to be a paradoxical result in view of the mathematical proofs, both in weak \cite{BCT06,BC03,BCG08,BCG09,BCT03,BG06} and strong \cite{CCC09,FPTT10} senses, that the \emph{scaled} distribution function $f^*$ tends for long times toward  the \emph{isotropic} HCS {self-similar} solution $\phi_\infty$ for any initial state (isotropic or anisotropic) with finite second-degree moments. The solution of the paradox lies in the fact that the above convergence properties do not imply that \emph{any}
moment of $f^*$ of degree higher than two should converge toward the corresponding moment of $\phi_\s$. In fact, our results provide a counter-example of that strong moment-based convergence property. On the other hand, we show that the ratio between an anisotropic moment and the isotropic moment of the same degree goes to zero.}

\section{The inelastic Maxwell model}

In the absence of external forces, the inelastic Boltzmann equation for a granular gas reads \cite{BP04}
 \beq
\left(\partial_t +\mathbf{v}\cdot\nabla\right)f(\mathbf{r},\mathbf{v};t)=J[\mathbf{v}|f,f], \label{2.1}
\eeq
where $J[\mathbf{v}|f,f]$ is the Boltzmann collision operator. The form of the operator $J$ for the IMM can be
obtained from the form for IHS by replacing the IHS collision frequency (which is proportional to the relative
velocity of the two colliding particles) by an effective velocity-independent collision frequency \cite{BK03}. With this
simplification, the velocity integral of the product $h(\mathbf{v})J[\mathbf{v}|f,f]$, where $h(\mathbf{v})$ is an arbitrary test function (``weak'' form of $J$), becomes
\beqa
\label{5J}
\int \dd{\bf v}_1 h({\bf v}_1)J[{\bf
v}_1|f,f]&=&\frac{\omeganu}{n\Omega_d} \int \dd{\bf v}_{1}\,\int
\dd{\bf v}_{2}
f({\bf v}_{1})f({\bf v}_{2})
\nn&&\times\int
\dd\widehat{\bm{\sigma}}\,\left[h({\bf v}_1'')-h({\bf
v}_1)\right],
\eeqa
where
\begin{equation}
\label{6}
{\bf v}_{1}''={\bf v}_{1}-\frac{1}{2}(1+\alpha)(
\widehat{\bm{\sigma}}\cdot {\bf
g})\widehat{\bm{\sigma}}
\end{equation}
denotes the post-collisional velocity, ${\bf g}={\bf v}_1-{\bf v}_2$ being the relative velocity and $\alpha\leq 1$ being
the constant coefficient of restitution, $n$ is the number density,  $\Omega_d=2\pi^{d/2}/\Gamma(d/2)$ is
the total solid angle in $d$ dimensions, and $\omeganu$ is the effective collision
frequency, which can be seen as a free parameter in the model. In particular, in order to get the same expression for the cooling rate as the one found for IHS
(evaluated in the local equilibrium approximation)  the adequate choice is \cite{BGM10,S03}
\begin{equation}
\label{4}
\omeganu=\frac{d+2}{2}\nu_0,\quad \nu_0=\frac{4\Omega_d}{\sqrt{\pi}(d+2)}n\sigma^{d-1}\sqrt{\frac{T}{m}},
\end{equation}
where $\sigma$ is the diameter of the spheres, $m$ is the mass, and $T$ is the granular temperature. However,  the results derived in this paper will be independent of the
specific choice of $\nu_0$.

In the case of Maxwell models (both elastic and inelastic), it is convenient to introduce the Ikenberry
polynomials \cite{TM80} $Y_{2r|i_1i_2\ldots i_s}(\mathbf{V})=V^{2r}Y_{i_1i_2\ldots
i_s}(\mathbf{V})$ of degree $k=2r+s$, where
$\mathbf{V}=\mathbf{v}-\mathbf{u}(\mathbf{r})$ is the peculiar velocity, $\mathbf{u}(\mathbf{r})$ being the mean
flow velocity. The $s$th-degree polynomials $Y_{i_1i_2\ldots i_s}(\mathbf{V})$ are obtained by subtracting from
$V_{i_1}V_{i_2}\ldots V_{i_s}$ that homogeneous symmetric polynomial of degree $s$ such as to make $Y_{i_1i_2\ldots i_s}(\mathbf{V})$ vanish upon contraction  on
any pair of indices. In particular, for $s=0$, 1, and 2 one has
\beq
Y_{2r|0}(\mathbf{V})=V^{2r},\quad Y_{2r|i}(\mathbf{V})=V^{2r}V_i, \label{X0}
\eeq
\beq
Y_{2r|ij}(\mathbf{V})=V^{2r}\left(V_iV_j-\frac{1}{d}V^2\delta_{ij}\right). \label{X1}
\eeq
Henceforth we will use the notation $M_{2r|\bar{s}}$ and $J_{2r|\bar{s}}$, where   $\bar{s}\equiv i_1i_2\ldots i_s$, for the moments and collisional moments, respectively, associated with the polynomials $Y_{2r|\bar{s}}(\mathbf{V})$. Note that the collisional moments are defined by Eq.\ \eqref{5J} with $h\to Y_{2r|\bar{s}}$.

As said before, the mathematical structure of the Maxwell collision
operator  implies that a collisional moment of degree $k$
can be expressed in terms of velocity moments of a degree less than
or equal to $k$. More specifically,
\beq
J_{2r|\bar{s}}=-\nu_{2r|s}
M_{2r|\bar{s}}+\sum_{r',r'',\bar{s}',\bar{s}''}^\dagger\lambda_{r'r''|\bar{s}'\bar{s}''\bar{s}}M_{2r'|\bar{s}'}M_{2r''|\bar{s}''},
\label{2.4}
\eeq
where the dagger in the summation denotes the constraints
$2(r'+r'')+s'+s''=2r+s$, $2r'+s'\geq 2$, and $2r''+s''\geq 2$. Since
the first term on the right-hand side of  Eq.\ \eqref{2.4} is
linear, then $\nu_{2r|s}$ represents the \emph{collisional rate} associated with the polynomial
$Y_{2r|\bar{s}}(\mathbf{V})$.
In particular,
\beq
\nutz=\frac{d+2}{4d}\left(1-\al^2\right)\nu_0,
\label{X6a}
\eeq
\beq
\nuzt=\frac{(1+\al)(d+1-\al)}{2d}\nu_0=\nutz+\frac{(1+\alpha)^2}{4}\nu_0.
\label{X6}
\eeq
The quantity $\nutz$ is actually
the \emph{cooling rate}, i.e., the rate of change of the granular
temperature due to the inelasticity of collisions. In general, it is possible to decompose  $\nu_{2r|s}$ as
\beq
\nu_{2r|s}=\frac{2r+s}{2}\nutz+\omega_{2r|s}.
\label{omega}
\eeq
The first term is the one inherent to the collisional cooling, while the second term ($\omega_{2r|s}$) can be seen as a \emph{shifted} collisional rate associated with the
 \emph{scaled} moment
\beq
M_{2r|\bar{s}}^*\equiv \frac{M_{2r|\bar{s}}}{n(2T/m)^{(2r+s)/2}}.
\eeq

The explicit forms for the collisional rates $\nu_{2r|s}$ and the $\lambda$ coefficients appearing in Eq.\ \eqref{2.4} have been evaluated in Ref.\ \cite{GS07} for $2r+s\leq 4$ and general $d$.

\section{Evaluation of $\nu_{2r|0}$, $\nu_{2r|1}$, and $\nu_{2r|2}$}
The aim of this section is to evaluate the collisional rates $\nu_{2r|0}$, $\nu_{2r|1}$, and $\nu_{2r|2}$ associated with the polynomials \eqref{X0} and \eqref{X1} as functions of the coefficient of restitution and the dimensionality. The procedure consists of inserting the polynomials $h=Y_{2r|0}$, $h=Y_{2r|i}$, and $h=Y_{2r|ij}$ into Eq.\ \eqref{5J} and \emph{focusing} only on the term proportional to the moments $M_{2r|0}$, $M_{2r|i}$, and $M_{2r|ij}$, respectively.

Let us describe the method with some detail in the case of $\nu_{2r|0}$. {}From the collision rule \eqref{6} one gets
\beqa
{V_1''}^{2r}-{V_1}^{2r}&=&\sum_{\ell=1}^r\binom{r}{\ell}{V_1}^{2(r-\ell)}(1+\al)^{\ell}\sg^{\ell}\nn
&&\times\left[\frac{1+\al}{4}\sg-\sv\right]^{\ell}.
\label{3}
\eeqa
This equation expresses the difference ${V_1''}^{2r}-{V_1}^{2r}$ as a linear combination of terms of order $V_1^{r_1}V_2^{r_2}$ with $r_1+r_2=2r$. Now, we need to extract those terms of order $V_1^{2r}$ and $V_2^{2r}$ only. The terms of order $V_1^{2r}$ are obtained from Eq.\ \eqref{3} by formally replacing $\mathbf{g}\to \mathbf{V}_1$, while the terms of order $V_2^{2r}$ are obtained  by formally replacing $\mathbf{g}\to -\mathbf{V}_2$ and taking the term corresponding to $\ell=r$ in the summation. Therefore,
\beqa
{V_1''}^{2r}-{V_1}^{2r}
&=&\sum_{\ell=1}^r\binom{r}{\ell}{V_1}^{2(r-\ell)}(1+\al)^{\ell}\left(\frac{\alpha-3}{4}\right)^{\ell}
\sv^{2\ell}\nn &&+\atwop^{2r}\svv^{2r}+\Delta_{2r|0}(\mathbf{V}_1,\mathbf{V}_2),
\label{3b}
\eeqa
where $\Delta_{2r|0}(\mathbf{V}_1,\mathbf{V}_2)$ denotes terms of order $V_1^{r_1}V_2^{r_2}$ with $r_1+r_2=2r$, $r_1\neq 0$, and $r_2\neq 0$. When inserting Eq.\ \eqref{3b} into Eq.\ \eqref{5J}, and ignoring $\Delta_{2r|0}(\mathbf{V}_1,\mathbf{V}_2)$, we obtain $-\nu_{2r|0}M_{2r|0}$ with the following expression for $\nu_{2r|0}$:
\beqa
\nu_{2r|0}&=&-\frac{\omeganu}{\Omega_d}\left[
\sum_{\ell=1}^r\binom{r}{\ell}(1+\al)^{\ell}\left(\frac{\alpha-3}{4}\right)^{\ell}B_{\ell}\right.\nn
&&\left.+\atwop^{2r}B_r\right],
\label{4k}
\eeqa
where
$B_\ell\equiv \int
\dd\widehat{\bm{\sigma}}\,(\widehat{\bm{\sigma}}\cdot
{\widehat{\bf
g}})^{2\ell}=2\pi^{(d-1)/2}
\Gamma\left(\ell+\frac{1}{2}\right)/{\Gamma\left(\ell+\frac{d}{2}\right)}$.
Equation \eqref{4k} can be rewritten in a more compact form as
\beq
\nu_{2r|0}=\nu_0\frac{d+2}{2}\left[1-\atwop^{2r}\frac{(\frac{1}{2})_r}{(\frac{d}{2})_r}-
{}_2\!F_1\left(-r,\frac{1}{2};\frac{d}{2};z\right)\right],
\label{16}
\eeq
where  $(a)_r$ denotes the Pochhammer symbol \cite{AS72}, $_2\!F_1(a,b;c;z)$ is the hypergeometric function \cite{AS72}, and $z\equiv (1+\al)(3-\al)/4$.
Equation \eqref{16} agrees with the result derived by Ernst and Brito \cite{EB02b} by a different method.

Proceeding in a similar way, and after lengthy algebra, one can evaluate the collisional rates $\nu_{2r|1}$ and $\nu_{2r|2}$. The results are
\beqa
\nu_{2r|1}&=&\nu_0\frac{d+2}{2}\left[1-\atwop^{2r+1}\frac{(\frac{3}{2})_r}{d(1+\frac{d}{2})_r}\right.
\nn
&&\left.-
{}_2\!F_1\left(-r,\frac{1}{2};\frac{d}{2};z\right)
+\frac{1+\al}{2d}\,{}_2\!F_1\left(-r,\frac{3}{2};\frac{d+2}{2};z\right)\right],\nn
\label{17}
\eeqa
\beqa
\nu_{2r|2}&=&\nu_0\frac{d+2}{2}\left[1-\atwop^{2(r+1)}\frac{r+1}{d(1+d/2)}\frac{(\frac{3}{2})_r}{(2+\frac{d}{2})_r}\right.\nn
&&-
{}_2\!F_1\left(-r,\frac{1}{2};\frac{d}{2};z\right)
+\frac{z}{d}\,{}_2\!F_1\left(-r,\frac{3}{2};\frac{d+2}{2};z\right)\nn
&&\left.+\atwop^2\frac{1}{2+d}\,{}_2\!F_1\left(-r,\frac{3}{2};\frac{d+4}{2};z\right)\right].
\label{18}
\eeqa
Note that, since $r$ is integer, the hypergeometric function $_2\!F_1(-r,b;c;z)$ is a polynomial in $z$ of degree $r$.

In the one-dimensional case ($d=1$), Eqs.\ \eqref{16} and \eqref{17} become
\beq
\nu_{2r|0}=\frac{3}{2}\nu_0\left[1-\atwop^{2r}-\left(\frac{1-\al}{2}\right)^{2r}\right],
\label{11}
\eeq
\beq
\nu_{2r|1}=\frac{3}{2}\nu_0\left[1-\atwop^{2r+1}-\left(\frac{1-\al}{2}\right)^{2r+1}\right].
\label{12}
\eeq
These expressions coincide with those previously derived in Ref.\ \cite{BK00}.

\begin{figure}
\begin{center}
\includegraphics[width=\columnwidth]{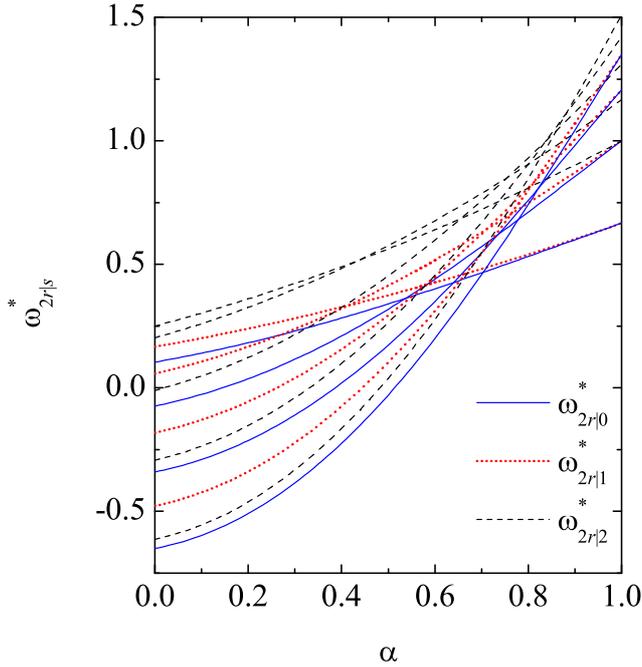}
\end{center}
\caption{Plot of (from  bottom to top at $\alpha=0$)
$\omega_{10|0}^*$, $\omega_{8|2}^*$, $\omega_{8|1}^*$, $\omega_{8|0}^*$,
$\omega_{6|2}^*$, $\omega_{6|1}^*$, $\omega_{6|0}^*$, $\omega_{4|2}^*$
$\omega_{4|1}^*$, $\omega_{4|0}^*$, $\omega_{2|1}*$,  $\omega_{2|2}^*$,  and $\omega_{0|2}^*$.
The dimensionality is $d=3$.
\label{fig1}}
\end{figure}

Figure \ref{fig1} displays the $\alpha$-dependence of the (scaled) shifted collisional rates $\omega_{2r|s}^*\equiv\omega_{2r|s}/\nu_0$ with $s=0,1,2$ and $2r+s\leq 10$ for the three-dimensional case ($d=3$). Of course, the null collisional rates $\omega_{0|0}=\omega_{0|1}=\omega_{2|0}=0$ are not plotted.
Several comments are in order. Firstly, the degeneracy
$\omega_{2r-2|1}=\omega_{2r|0}$ present in the elastic limit \cite{S09b,TM80} is broken,
yielding $\omega_{2r-2|1}<\omega_{2r|0}$. Analogously, the linear relationship $d\omega_{2r|1}=(d-1)\omega_{2r-2|2}+\omega_{2r|0}$ for  elastic Maxwell particles no longer holds if $\alpha<1$, except in the case $r=1$, where one has $d\omega_{2|1}=(d-1)\omega_{0|2}$ for any  $\alpha$ \cite{GS07}.
Secondly, we observe that all the shifted collisional rates monotonically decrease with increasing dissipation, eventually becoming negative, except those corresponding to $2r+s\leq 5$. The physical implications of this change of sign will be discussed in the next section.
A further observation that can be extracted from Fig.\ \ref{fig1} is that the impact of $\alpha$ on $\omega_{2r|s}$ becomes generally more pronounced as the degree $2r+s$ increases.
In the case of the unshifted collisional rates $\nu_{2r|s}$, a graph similar to Fig.\ \ref{fig1} (not reported here) shows a non-monotonic dependence on $\alpha$: they first increase with increasing inelasticity, reach a maximum, and then decrease smoothly. In contrast to the shifted collisional rates $\omega_{2r|s}$, the collisional rates $\nu_{2r|s}$ are always positive, as expected on physical grounds.

\section{Diverging moments in the HCS}

The Boltzmann equation for the HCS is given by Eq.\ \eqref{2.1} with $\nabla\to 0$. It is more convenient to rewrite it in terms of the \emph{scaled} distribution
\beq
f^*(\mathbf{c}(t),t)=\frac{1}{n}[{2T(t)}/{m}]^{d/2}f(\mathbf{v},t) ,
\quad \mathbf{c}(t)={\mathbf{v}}/{\sqrt{2T(t)/m}}.
\label{3.1}
\eeq
The resulting Boltzmann equation is
\beq
\partial_\tau f^*(\mathbf{c},\tau)+\frac{\nu_{0|2}^*}{2}\frac{\partial}{\partial{\mathbf{c}}}\cdot\left[\mathbf{c}f^*(\mathbf{c},\tau)\right]=J^*[\mathbf{c}|f^*,f^*],
\label{4.1}
\eeq
where $\dd\tau=\nu_0\dd t$, $\nu_{2|0}^*\equiv \nu_{2|0}/\nu_0$ is the reduced cooling rate, and $J^*$ is the dimensionless Boltzmann collision operator.
{}From Eq.\ \eqref{4.1}, and taking into account Eq.\ \eqref{2.4}, one gets the time evolution equation of the moments:
\beq
\partial_\tau M_{2r|\bar{s}}^*=-\omega_{2r|s}^*
M_{2r|\bar{s}}^*+\frac{n}{\nu_0}\sum_{r',r'',\bar{s}',\bar{s}''}^\dagger\lambda_{r'r''|\bar{s}'\bar{s}''\bar{s}}M_{2r'|\bar{s}'}^*M_{2r''|\bar{s}''}^*.
\label{4.2}
\eeq

If the distribution function is {isotropic}, i.e., $f^*(\mathbf{c},\tau)=f^*(c,\tau)$, then the only non-vanishing moments are $M_{2r|0}^*(\tau)$. We will refer to them as the \emph{isotropic} moments. On the other hand, if the initial distribution function $f^*(\mathbf{c},0)$ is not isotropic, the other moments, in particular $M_{2r|i}^*(\tau)$ and $M_{2r|ij}^*(\tau)$, are not necessarily zero. We will call \emph{anisotropic odd} moments to $M_{2r|i}^*(\tau)$ and \emph{anisotropic even} moments to $M_{2r|ij}^*(\tau)$.

Since the time evolution of the \emph{scaled} velocity moments in
the HCS is governed by the shifted collisional rates
$\omega_{2r|s}$, the fact that the latter can become negative (for $\alpha$ smaller than a certain threshold value depending on $r$ and $s$) implies
that the associated moments diverge in time.

Among the (scaled)
moments $M_{2r|0}^*$, $M_{2r|i}^*$, and $M_{2r|ij}^*$, Fig.\
\ref{fig1} shows that the lowest-degree diverging moments are
(in the three-dimensional case)  the sixth-degree moments $M_{4|ij}^*$ and $M_{6|0}^*$, which diverge for $\alpha\leq 0.020$ and $\alpha\leq 0.145$, respectively.
Moments of  higher degree diverge for smaller inelasticities. More specifically,
 $M_{6|i}^*$, $M_{6|ij}^*$, $M_{8|0}^*$, $M_{8|i}^*$, $M_{8|ij}^*$, and $M_{10|0}^*$ diverge for $\alpha$ smaller than  $0.261$, $0.331$, $0.386$, $0.444$, $0.482$, and $0.514$, respectively. In general, the larger the degree the larger the threshold value of the coefficient of restitution below which the moment diverges. Given a degree $2r$, the isotropic moment $M_{2r|0}^*$ diverges earlier (i.e., with a larger threshold value $\alpha=\alpha_{2r|0}$) than the anisotropic (even) moment $M_{2r-2|ij}^*$.
The threshold value of $\alpha_{2r|0}$ can be obtained as the solution of the equation
$\omega_{2r|0}=0$. {}From Eq.\ \eqref{16}, this is
equivalent to
\beq
\frac{r}{2d}(1-\al^2)=1-\atwop^{2r}\frac{(\frac{1}{2})_r}{(\frac{d}{2})_r}-
{}_2\!F_1\left(-r,\frac{1}{2};\frac{d}{2};z\right).
\label{19}
\eeq
Given an integer value of $r$, Eq.\ \eqref{19} is an equation of
degree $2r$ in $\alpha$.

The Boltzmann equation \eqref{4.1} for the scaled distribution function $f^*(\mathbf{c},\tau)$ admits a stationary and \emph{isotropic} solution $\phi_{\s}(c)$. This corresponds to a {\emph{self-similar}} solution to the original Boltzmann equation where all the velocity and time dependence is encapsulated in the scaled velocity $\mathbf{c}$.
{About ten years ago, Ernst and Brito \cite{EB02a,EB02b} conjectured that the general solution of Eq.\ \eqref{4.1}  asymptotically tends to  $\phi_{\s}(c)$ for long times. Let us loosely express this conjecture as}
\beq
\lim_{\tau\to\infty}f^*(\mathbf{c},\tau)=\phi_{\s}(c),
\label{4.3}
\eeq
{where the precise meaning of the limit needs to be fixed in a rigorous mathematical sense. The existence of the self-similar  solution and the convergence rate for the general approach to
this state was first addressed  in Ref.\ \cite{BC03}. However, in that
work the authors imposed conditions that were proven to be unnecessary in Refs.\ \cite{BCG08,BCG09,BG06}. More recently, proofs of the strong convergence in Sobolev and $L^1$ norms for small \cite{CCC09} and finite \cite{FPTT10} inelasticity have been published. Those proofs hold for any initial data (probability densities with bounded second-degree moments), regardless of being isotropic or not, but they do not imply that any moment of degree higher than two converges to the corresponding moment of the self-similar solution. This stronger moment-to-moment interpretation of the Ernst--Brito conjecture would read}
\beq
{\lim_{\tau\to\infty}M_{2r|\bar{s}}^*(\tau)=\int \dd\mathbf{c}\, Y_{2r|\bar{s}}(\mathbf{c})\phi_{\s}(c).}
\label{EBnew}
\eeq
{As discussed below, this  stronger notion of the convergence statement \eqref{4.3} does not hold.}

Although the explicit form of $\phi_{\s}(c)$ is not known, except in the one-dimensional case \cite{BMP02}, it is known that it possesses an algebraic high-velocity tail of the form $\phi_{\s}(\mathbf{c})\sim c^{-d-\gamma_0(\alpha)}$, where $\gamma_0(\alpha)$ obeys  a transcendental equation {\cite{BK02,BCG08,BCG09,EB02a,EB02b,KB02}}. As a consequence, the {isotropic} moments $M_{2r|0}^*$ with $2r\geq \gamma_0(\alpha)$ diverge. {According to the strong  convergence property \eqref{EBnew}, this would imply} that, if  $ M_{2r|0}^*(0)=\text{finite}$, then $\lim_{\tau\to\infty} M_{2r|0}^*(\tau)=\infty$ if $2r\geq \gamma_0(\alpha)$. This is fully consistent with the fact that $\omega_{2r|0}<0$, so that $M_{2r|0}^*(\tau)$ {indeed} diverges in time if $\alpha<\alpha_{2r|0}$. In fact, formally setting $2r=\gamma_0$ in
Eq.\ \eqref{19} one recovers the transcendental equation for
$\gamma_0$ derived by an independent method \cite{BK02,EB02a,EB02b,KB02}.

The interesting point is
that, as shown above, the \emph{anisotropic} moments $M_{2r|i}^*$ and $M_{2r|ij}^*$
can also diverge, unless they are zero in the initial state. {The possibility that $\lim_{\tau\to\infty} M_{2r|i}^*(\tau)=\infty$ and $\lim_{\tau\to\infty} M_{2r|ij}^*(\tau)=\infty$ contradicts the strict moment-to-moment limit \eqref{EBnew}}, since all  the anisotropic moments of $\phi_{\s}(c)$ vanish. Let us elaborate this {result} in more detail.

In principle, we have derived Eqs.\ \eqref{16}--\eqref{18} for
$r=\text{integer}$. However, since the hypergeometric function and
the Pochhammer symbols are well defined for $r\neq\text{integer}$, {we speculate that
an analytic continuation of Eqs.\
\eqref{16}--\eqref{18} to $r\neq\text{integer}$ is possible.} It is then tempting to
interpret $\omega_{k|0}^*$, $\omega_{k-1|1}^*$, and $\omega_{k-2|2}^*$ as the quantities governing  the asymptotic
time evolution of the averages $M_{k|0}^*\equiv\langle c^k\rangle$, $M_{k-1|i}^*\equiv\langle c^{k-1}c_i\rangle$, and $M_{k-2|ij}^*\equiv\langle c^{k-2}\left(c_ic_j-d^{-1}c^2\delta_{ij}\right)\rangle$, respectively, even if
$k/2\neq\text{integer}$ and $(k-1)/2\neq\text{integer}$, {although a formal proof of this expectation is beyond the scope of this paper.}
As said before, $M_{k|0}^* \to\infty $ if $k>\gamma_0(\alpha)$, where $\omega_{\gamma_0|0}^*=0$. Analogously, we can expect that the anisotropic quantities $M_{k-1|i}^*$ and $M_{k-2|ij}^*$ diverge if $k>\gamma_1(\alpha)$ and $k>\gamma_2(\alpha)$, respectively, where $\gamma_1$ and $\gamma_2$ are the solutions to the equations $\omega_{\gamma_1-1|1}^*=0$ and $\omega_{\gamma_2-2|2}^*=0$.

\begin{figure}
\begin{center}
\includegraphics[width=\columnwidth]{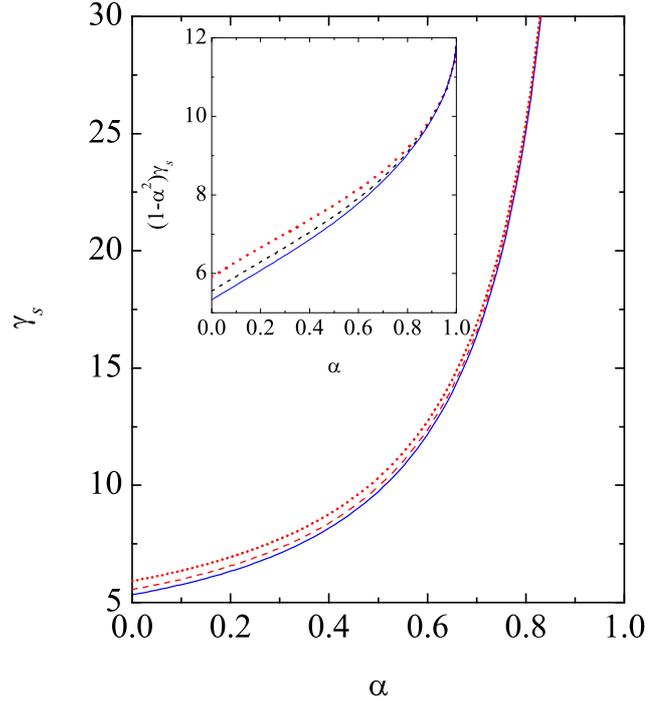}
\end{center}
\caption{Plot of (from bottom to top) $\gamma_0(\alpha)$,
$\gamma_1(\alpha)$, and $\gamma_2(\alpha)$. The inset shows $(1-\alpha^2)\gamma_s$ versus $\alpha$. The
dimensionality is $d=3$.
\label{fig2}}
\end{figure}

The functions $\gamma_0(\alpha)$,
$\gamma_1(\alpha)$, and $\gamma_2(\alpha)$ are
displayed in Fig.\ \ref{fig2} for $d=3$. In the elastic limit $\alpha\to 1$, the three exponents diverge as $\gamma_s\approx 4d/(1-\alpha^2)$ \cite{BK02,KB02}, as shown in the inset of Fig.\ \ref{fig2}.
We observe that
$\gamma_0(\alpha)<\gamma_1(\alpha)<\gamma_2(\alpha)$. This
implies that, at a given value of $\alpha$ the isotropic average $M_{k|0}^*$ starts to diverge before the anisotropic (odd) average  $M_{k-1|i}^*$ does, and the latter does it before the anisotropic (even) average $M_{k-2|ij}^*$ does. Stated differently, if we focus on the ratios between the
anisotropic and the isotropic averages, we can expect the asymptotic behaviors
\beq
\frac{M_{k-1|i}^*}{M_{k|0}^*}\sim
 e^{-(\omega_{k-1|1}^*-\omega_{k|0}^*)\tau},\quad \frac{M_{k-2|ij}^*}{M_{k|0}^*}\sim
e^{-(\omega_{k-2|2}^*-\omega_{k|0}^*)\tau}.
\label{23}
\eeq
Since $\omega_{k-2|2}^*>\omega_{k-1|1}^*>\omega_{k|0}^*$, it turns out that
\beq
\lim_{\tau\to\infty}\frac{M_{k-1|i}^*}{M_{k|0}^*}=0,\quad \lim_{\tau\to\infty}\frac{M_{k-2|ij}^*}{M_{k|0}^*}=0.
\label{25}
\eeq
Therefore, the anisotropic moments, \emph{relative to the isotropic moments of the same degree}, asymptotically go to zero (the anisotropic even moments more rapidly than the anisotropic odd ones). {}From that point of view, Eq.\ \eqref{25}  can be seen as a {\emph{weak} validation of a moment-to-moment interpretation of Eq.\ \eqref{4.3} for initial anisotropic distributions.}

The one-dimensional system deserves some separate comments. In that case, the {self-similar} solution is $\phi_\s(c)=(2^{3/2}/\pi)(1+2c^2)^{-2}$ \cite{BMP02}, so that $\gamma_0=3$ and the moments $\langle c^k\rangle $ with $k\geq 3$ diverge. This agrees with Eq.\ \eqref{11}, according to which $\omega_{k|0}^*\leq 0$ for $k\geq 3$.
Analogously, from Eq.\
\eqref{12} one gets $\gamma_1=3$.
In particular, the isotropic moment $\langle c^3\rangle$ diverges, while the anisotropic moment  $\langle c^2 c_x\rangle$ (proportional to the heat flux) keeps its initial value \cite{BK00,GS07}. Therefore, $\langle c^2 c_x\rangle/\langle c^3\rangle\to 0$. On the other hand, since $\omega_{k|0}^*=\omega_{k-1|1}^*<0$ for $k>3$, there exist two possible scenarios for the ratios $\langle c^{k-1} c_x\rangle/\langle c^k\rangle$: either they tend to constant values or they decay more slowly than exponentially. A deeper investigation is needed to elucidate between these two possibilities.

\section{Conclusion}
To summarize, we have shown that the {strong notion \eqref{EBnew} of the Ernst--Brito conjecture cannot be strictly true since it does not hold for anisotropic initial conditions. However, we \emph{conjecture} that $M^*_{2r|\bar{s}}(\tau)/M^*_{2r+s|0}(\tau)\to 0$ when $\bar{s}\neq 0$, even if  $M^*_{2r|\bar{s}}(\tau)\to \infty$, as shown by Eq.\ \eqref{25} for $\bar{s}=i$ ($s=1$) and $\bar{s}=ij$ ($s=2$). In order to elaborate further this conjecture}, let us decompose $f^*(\mathbf{c},\tau)$ into its isotropic, anisotropic symmetric, and
antisymmetric parts:
\beq
f^*(\mathbf{c},\tau)=\phi(c,\tau)+\widetilde{f}_+^*(\mathbf{c},\tau)+f_-^*(\mathbf{c},\tau),\quad
\label{27}
\eeq
where
\beq
\widetilde{f}_+^*(\mathbf{c},\tau)\equiv f_+^*(\mathbf{c},\tau)-\phi({c},\tau),\quad
\phi({c},\tau)\equiv \frac{1}{\Omega_d}\int\dd
\widehat{\mathbf{c}}\, f_+(\mathbf{c},\tau),
\label{30}
\eeq
\beq
f_\pm^*(\mathbf{c},\tau)\equiv
\frac{1}{2}\left[f^*(\mathbf{c},\tau)\pm
f^*(-\mathbf{c},\tau)\right].
\eeq
As a consequence, the velocity moments $M_{k|0}^*(\tau)$, $M_{k-1|i}^*(\tau)$, and $M_{k-2|ij}^*(\tau)$ are related to $\phi(c,\tau)$, ${f}_-^*(\mathbf{c},\tau)$, and $\widetilde{f}_+^*(\mathbf{c},\tau)$, respectively.
If
the ``sizes'' of these three contributions are
measured through those three classes of moments, we can say that, as time progresses, the two anisotropic  parts of $f^*$
become negligible versus the isotropic part, i.e.,
$|f_-^*(\mathbf{c},\tau)|\ll \phi({c},\tau)$ and
$|\widetilde{f}_+^*(\mathbf{c},\tau)|\ll \phi({c},\tau)$,  \emph{in
the sense} of Eq.\ \eqref{25}. Moreover, $\lim_{\tau\to\infty}\phi(c)=\phi_{\s}(c)$.
We further speculate that the \emph{high-velocity tails} of the anisotropic contributions tend to the forms
\beq
f_-^*(\mathbf{c},\tau)\to \chi_-(\widehat{\mathbf{c}})c^{-d-\gamma_1(\alpha)}, \quad \widetilde{f}_+^*(\mathbf{c},\tau)\to \widetilde{\chi}_+(\widehat{\mathbf{c}})c^{-d-\gamma_2(\alpha)},
\eeq
where the angular functions $\chi_-(\widehat{\mathbf{c}})$ and $\widetilde{\chi}_+(\widehat{\mathbf{c}})$ depend on the initial conditions.
A confirmation of the above expectations requires a more refined analysis.

\begin{acknowledgements}
This paper is dedicated to the memory of  Isaac Goldhirsch, who was always a source of inspiration, scientific integrity, and human quality.
{The authors are grateful to the two anonymous referees for their  comments and suggestions.}
Support from  the Ministerio de  Ciencia e Innovaci\'on (Spain) through Grant No.\ {FIS2010-16587} and  from the Junta de Extremadura (Spain) through Grant No.\ GR10158 (partially financed by FEDER funds) is  gratefully acknowledged.
\end{acknowledgements}

\bibliographystyle{spmpsci}      % mathematics and physical sciences

\bibliography{D:/bib_files/Granular}
\end{document}